\documentclass[%
reprint,
showkeys,
nofootinbib,
amsmath,amssymb,
aps,
prl,
]{revtex4-2}
\usepackage[marginal]{footmisc}
\usepackage{graphicx}
\usepackage{dcolumn}
\usepackage{bm}
\usepackage[colorlinks,allcolors=black,CJKbookmarks=True]{hyperref}

\begin{document}
	
\title{Phase transition catalyzed by primordial black holes}

\author{Zhen-Min Zeng$^{1,2}$ }
\email{cengzhenmin@itp.ac.cn}

\author{Zong-Kuan Guo$^{1,2,3}$}
\email{guozk@itp.ac.cn}

\affiliation{$^{1}$CAS Key Laboratory of Theoretical Physics, Institute of Theoretical Physics, Chinese Academy of Sciences, Beijing 100190, China }
\affiliation{$^{2}$School of Physical Sciences, University of Chinese Academy of Sciences, Beijing 100049, China }
\affiliation{$^{3}$School of Fundamental Physics and Mathematical Sciences, Hangzhou Institute for Advanced Study, University of Chinese Academy of Sciences, Hangzhou 310024, China}

\begin{abstract}
We investigate the first-order phase transition catalyzed by primordial black holes~(PBHs) in the early Universe.
We find that super-horizon curvature perturbations generated in this scenario
lead to the production of gravitational waves when the scalar modes re-enter the horizon.
If PBHs with masses about $10^{-13}M_{\odot}$ constitute all dark matter,
the first-order electroweak phase transition catalyzed by PBHs can explain the gravitational wave signal
observed by pulsar timing array collaborations without the overproduction of PBHs.
\end{abstract}
\maketitle

\emph{Introduction.}
First-order phase transitions~(PTs) are predicted by new physics beyond the standard model~\cite{PhysRevD.55.3873,PhysRevD.56.2893,BODEKER1997387,LAINE199643}.
This process can generate gravitational waves~(GWs)
through bubble collisions~\cite{PhysRevD.49.2837,Huber_2008}, sound waves~\cite{PhysRevLett.112.041301,PhysRevD.96.103520,PhysRevD.92.123009}
and fluid turbulence~\cite{PhysRevD.74.063521,ChiaraCaprini_2009,PhysRevD.66.024030,PhysRevD.76.083002,Peter_2018}.
Recent observations of the stochastic gravitational-wave backgrounds (SGWBs) from pulsar timing array collaborations
implies a supercooling quantum chromodynamics PT at low energy scales
if the signal is generated by bubble collisions~\cite{ellis_what_2023,afzal_nanograv_2023,agazie_nanograv_2023},
which infers a finite baryon chemical potentials. 

These PTs are assumed to take place in a flat space-time.
The effect of local strong gravitational fields~(such as PBHs) are not considered in the evolution of the first-order PT.
Recent studies indicate that PBHs can significantly enhance the vacuum decay rate~\cite{hiscock_can_1987,PhysRevD.32.1333,gregory_black_2014,burda_vacuum_2015,burda_fate_2016,Canko:2017ebb,gregory_black_2020,PhysRevD.101.125012,gregory_hawking-moss_2020,jinno_super-slow_2023,mukaida_false_2017,kohri_electroweak_2018,el-menoufi_black_2020,hayashi_catalyzed_2020,hamaide_primordial_2023}. The enhancement effect includes two parts,
the change of the Bekenstein entropy and the decrement of the action
that depends on the Euclidean bubble wall trajectory.
The former dominates when PBH masses are of order of the plank mass.
It is under debate that the thermal effect of such small PBHs is so strong
that may stabilize the fields in the symmetric phase~\cite{strumia_black_2023}.
The latter dominates when PBHs are large,
especially when the horizon radius of PBHs is comparable to the bubble 
radius~\cite{hiscock_can_1987,OSHITA2019149}.
In this case the bubble nucleation rate is strongly enhanced due to the strong gravitational effect.


In this Letter, we consider the catalyzing effect of PBHs on a thermal first-order PT.
We assume that the vacuum energy density is vanishingly small in the true vacuum at low temperature.
Due to the catalyzing effect of PBHs,
the tunneling process occurs before the temperature decreases to the PT temperature, $T_n$,
which is determined by $\Gamma(T_n)\sim H^4(T_n)$~\cite{dine_towards_1992,ellis_maximal_2019,cai_gravitational-wave_2017}.
It reveals a tunneling process from a de Sitter vacuum to a de Sitter vacuum with a relatively small vacuum energy.
This differs from the thermal PT,
where the scalar field tunnels from a de Sitter vaccum to a Minkovski vacuum.

Assuming PBHs are sparse enough that the average number of PBHs in a Hubble-size region during PT is less than unity,
which indicates there are some Hubble-size regions that contain a PBH while the others don't.
In an identification Hubble-sized region that contains a PBH,
slightly after the degenerate vacuum forms at the critical temperature, $T_c$,
PT is triggered by the PBH and itself as a nucleation site.
However, the Hubble-size regions without PBHs stay in the false vacuum
until the nucleation rate $\Gamma(T)$ gets close to $H^4(T)$,
and then the nucleation of thermal true vacuum bubbles starts.
Such an asynchronism of vacuum decay leads to the different evolution of the probability of the false vacuum
and then affects the averaged equation of state in the different Hubble horizons during the PT,
finally induces curvature perturbations at super-horizon scales.
Since the asynchronism originates from the distribution of PBHs as catalysts,
induced density perturbations inherit the distribution of PBHs at super-horizon scales.

We analytically calculate the power spectrum of curvature perturbations by deriving the number density of PBHs.
We find that at scales larger than the average separation of PBHs,
the probability distribution of density perturbations is close to a finite-width Gaussian distribution.
The existence of the upper limit of density perturbations indicates
that this mechanism can generate a strong SGWB signal and avoid the overproduction of PBHs.
We apply our formalism to the electroweak phase transition~(EWPT).
If PBHs with masses $M\sim 10^{-13}M_{\odot}$ constitute all dark matter as catalysts~\cite{carr_constraints_2021},
the predicted SGWB 
can explain the observed signal from recent pulsar timing array collaborations~\cite{afzal_nanograv_2023,agazie_nanograv_2023,ellis_what_2023}
and avoid the overproduction of PBHs.

\emph{First-order PT catalyzed by PBHs.}
We regard PBHs as the nucleation sites of true vacuum bubbles during a thermal first-order PT.
Using the technique developed in Refs.~\cite{hiscock_can_1987,mukaida_false_2017},
the upper limit of the mass of PBHs that can catalyze the first-order PT
is estimated to be $m_{pl}^3/\sqrt{V_{fv}}$\,
where $m_{pl}\equiv 1/\sqrt{8\pi G}$ is the plank mass
and $V_{fv}$ is the energy density of the false vacuum.
This estimation ignores the change of the Bekenstein entropy,
which only considers the pure gravitational effect.
For the EWPT, $V_{fv}\sim 10^7 GeV^4$, which suggests
that PBHs with masses less than $10^{-4}M_{\odot}$ can catalyze the PT. If the PT is slow, $T_n$ is much smaller than $T_c$. The reduction of Euclidean action resulting from the catalyzing effect will greatly advance the bubble nucleation time, which suggests $T_b\gg T_n$, where $T_b$ is the formation temperature of bubbles catalyzed by PBHs. This is the case that we focus mainly on throughout this letter.

In a general first-order PT, the bubble nucleation rate of the true vacuum can be generally expressed in the exponential form~\cite{PhysRevD.16.1248,PhysRevD.45.3415,ellis_maximal_2019}
\begin{equation}
    \Gamma(t)=\Gamma_0 e^{\beta t},
    \label{eqa:dr}
\end{equation}
where $\Gamma_0$ and $\beta$ are approximately constant.
The parameter $\beta$ represents the increasing rate of vacuum bubbles.
In the case of $\beta\gg 1$ , $\beta^{-1}$ is approximately the PT duration.
The spatial fraction of the false vacuum in the thermal PT is~\cite{PhysRevD.46.2384}
\begin{equation}
    F(t)=\exp{\left[-\frac{4\pi}{3}\int_{t_i}^{t}dt'\Gamma(t')a^3(t')r^3(t',t)\right]},
    \label{eqa:ft}
\end{equation}
where $r(t',t)\equiv\int_{t'}^{t}\frac{1}{a(t'')}dt''$ denotes the comoving distance of true vacuum bubbles nucleated at $t'$,
and $t_i$ is the nucleation time of the first bubble.

However, in the case of the catalyzed PT,
the formula~\eqref{eqa:ft} for $F(t)$ is no longer valid.
The true vacuum bubbles first nucleate around the PBH at $T_b$
which is slightly lower than the critical temperature $T_c$.
When the temperature decreases to the PT temperature $T_n$ determined by $\Gamma(T_n)\sim H^{4}(T_n)$,
the thermal bubbles start to nucleate.
At this moment the false vacuum fraction is no longer 1 due to the expansion of the catalyzed bubbles.
The expression of $F(t)$ is modified to
\begin{equation}
    \frac{d F(t)}{dt}=-F(t)\int_{t_i}^{t}\Gamma(t')a^{3}(t')d t'\frac{d\Bar{V}_{tb}}{dt}-\frac{d\Bar{V}_{cb}}{\Bar{V_H}}dt,
    \label{eqa:ft_b}
\end{equation}
where $\Bar{V}_{tb}=\frac{4\pi}{3}(\int_{t_n}^{t}\frac{v_{w}(\Tilde{t})}{a(\Tilde{t})}d\Tilde{t})^3$
and $\Bar{V}_{cb}=\frac{4\pi}{3}(\int_{t_c}^{t}\frac{v_{w}(\Tilde{t})}{a(\Tilde{t})}d\Tilde{t})^3$
are the comoving volumes of bubbles for the thermal PT and the catalyzed PT, respectively,
$\Bar{V_H}=\frac{4\pi}{3}(\frac{1}{\mathcal{H}(t)})^3$ is the comoving Hubble volume, $\mathcal{H}$ is the comoving Hubble parameter, $t_c$ is the time that corresponds to temperature $T_c$, and $v_w$ is the speed of bubble walls. Here we simply take the assumption $T_c\approx T_b$ as we focus on a catalyzed slow phase transition.

Before $t_c$, the field stays at the global minimum.
Slightly after the degenerate vacua appear, catalyzed bubbles form
and expand due to the increasing pressure between the vacua,
leading to the decrease of $F(t)$.
When the temperature decrease to $T_n$,
thermal true vacuum bubbles start to nucleate and the decreasing of $F(t)$ is accelarated.
As $F(t)$ decreases, the vacuum energy is transferred to bubble walls and plasma. Since we mainly focus on the case that PBHs are too sparse that their average distance is larger than the horizon.  These catalyzed and non-catalyzed regions are separate Robertson Walker universes with different evolution of $F(t)$, which is equivalent to having different equation of state among those Hubble horizons. To deal with asynchronous evolution on superhorizon scale, we follow the the opinion of gradient expansion~\cite{Lyth:2004gb}. When the scale of the imhomogenity $L$ is larger than the horizon $H^{-1}$, there exists an appropriate set of coordinates with RW like metric. This assumption is detailed discussed in the Supplemental Material. The Friedmann equation and equations of motion read~\cite{liu_primordial_2022,liu_constraining_2023,PhysRevD.91.084057,escriva_primordial_2023}
\begin{eqnarray}
   \Tilde{H}^2&=&\frac{\rho_r+\rho_w+\rho_v}{3}, \label{eqa:fri}\\
    \rho_v&=&F(t)\Delta V,\label{eqa:rhov_eom}\\
    \frac{d(\rho_r+\rho_w)}{dt}&+&4\Tilde{H}(\rho_r+\rho_w)=-\frac{d\rho_v}{dt},
   \label{eqa:rhor_eom}
\end{eqnarray}
where $\Tilde{H}$ is the local Hubble parameter,
$\rho_r$, $\rho_w$ and $\rho_v$ are the energy densities of background radiation, bubble walls and the false vacuum, respectively.
$\Delta V$ is the energy density difference of the true and false vacua at the tunneling time.
Here we have assumed that the velocity of bubble walls is closed to 1,
which means that bubble walls are regarded as ultra-relativistic matter that share the same equation of motion parameter of background plasma~\cite{vilenkin}.

\begin{figure}[t!]
    \includegraphics[width=2.7in]{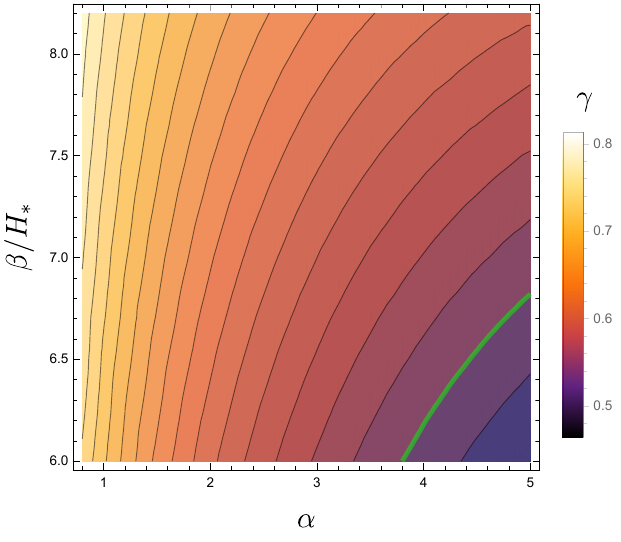}
    \caption{Energy density ratio $\gamma$ as a function of the PT parameters $\alpha$ and $\beta/H_*$ in the case of $T_b\sim T_c\gg T_n$.
    The green line corresponds to $\gamma=0.53$.}
    \label{fig:ga}
\end{figure}

The asynchronous PT scenario is firstly seen as the `late-blooming' mechanism~\cite{liu_primordial_2022,gouttenoire2023primordial}, where some regions may postpone the vacuum decay time. Since in the expanding Universe $\rho_r$ and $\rho_w$ decrease as $a^{-4}$ while $\Delta V$ remains almost constant, so if the false vacuum decays later, the delayed vacuum decay regions will have higher energy density after the PT.
On the contrary, the advanced vacuum decay results in the lower energy density,
which implies that the ratio of the energy density for the catalyzed PT to that for the thermal PT,
$\gamma = \rho_{cpt}/\rho_{tpt}$, is less than 1, as shown in Fig.~\ref{fig:ga}.
The values of $\rho_{cpt}$ and $\rho_{tpt}$ are straightforward calculated in a specific first-order PT model
that is described by $T_*$, $\alpha\equiv\Delta V/\rho_r(t_*)$ and $\beta/H_{*}$, where $T_*$ is evaluated at the percolation time $t_*$ with $F(t_*)=0.7$.
The disparities in energy between horizons result in curvature perturbations, from which we will deduce the statistical properties based on the distribution of PBHs. It is worth noting that this scenario is similar to the generation of the curvature perturbation from isocurvature initial perturbations.

\emph{Curvature perturbations from the catalyzed PT.}
In our scenario of the PT catalyzed by PBHs,
we assume that seed PBHs are monochromatic and form earlier than the PT.
The comoving number density of seed PBHs is $n_{\mathrm{PBH},c}(M)$.
The average number of PBHs with masses $M$ in a Hubble-size region at time $t$ is
\begin{equation}
    p(M,t)=n_{\mathrm{PBH},c}(M)\frac{4\pi}{3}\left(\frac{1}{\mathcal{H}(t)}\right)^3.
    \label{eqa:p}
\end{equation}
For monochromatic PBHs with masses $M$,
$p(M,t)$ increase as the Universe expands.
We are interested in the case of $p(M,t)<1$ during the PT.
In this case, PBHs randomly distribute in space
and satisfy the Poisson distribution in each comoving Hubble volume.
The probability distributions of density perturbations at the end of the PT
inherit the distributions of PBHs.
Now we investigate the statistical properties of curvature perturbations from the distribution of PBHs.

Consider a comoving volume of $(2L)^3$, where $L=n/\mathcal{H}(t_e)$, $n\gg1$ and $t_e$ is the end time of the PT. At that point, the Universe reheats and gets back to radiation-dominated epoch. To study the statistical properties of PBHs, $L$
should be much larger than the comoving mean separation,
$S=n_{\mathrm{PBH},c}^{1/3}(M)$, so that enough PBHs are
contained in the volume. Let $p$ denote the probability that a PBH presents in a Hubble volume
and $X_i$ denote the number of PBHs contained in the $i$th Hubble volume, where $i\leq n^3$ and $X_i=0$ or 1 by definition.
The expectation value and the varience of the random variable $X_{i}$ are respectively $E(X_i)=p$ and $D(X_i)=p(1-p)$.
Since the PBH number in the volume $V_L=(2L)^{3}$ is much larger than one,
according to the central limit theorem,
the total PBH number $X=\Sigma_{1}^{n^3}X_i$ in the comoving volume $(2L)^{3}$ is subject to the Gaussian distribution
with the expectation value $E(X)=n^3 E(X_i)$ and the variance $D(X)=n^3 D(X_{i})$.
Since $\rho_{cpt}$ and $\rho_{tpt}$ are the energy densities of the causality disconnect Hubble volumes
that contain or do not contain a PHB after the PT,
the energy density of a comoving volume $V_L$ is $\rho_X=[X\rho_{cpt}+(n^3-X)\rho_{tpt}]/n^3=\rho_{tpt}+\frac{X}{n^3}(\rho_{cpt}-\rho_{tpt})$.
$\rho_X$ is a stochastic variable that satisfies the Gaussian distribution.
Its expectation value and variance are $E(\rho_X)=\rho_{tpt}+p(\rho_{cpt}-\rho_{tpt})$ and $D(\rho_X)=(\rho_{cpt}-\rho_{tpt})^2p(1-p)/n^3$.
Then the density perturbation as a stochastic variable is written as
\begin{equation}
     \delta_{X}\equiv\frac{\rho_{X}-E(\rho_X)}{E(\rho_X)}=\frac{(\frac{X}{n^3}-p)(\gamma-1)}{1+p(\gamma-1)},
\end{equation}
where $\gamma<1$ by definition. We easily see that the density perturbation is still a Gaussian stochastic variable
and it is valid in the range of $\delta_X\in\left [\frac{(p-1)(1-\gamma)}{1+p(\gamma-1)},\frac{p(1-\gamma)}{1+p(\gamma-1)}\right ]$.
With the increasing of $n$, due to the central limit theorem,
the probability distribution of $\delta_X$ is getting closer and closer to the Gaussian distribution in that certain range.
It is known that, in radiation dominate epoch, when density perturbations exceed a threshold value $\delta_c\sim 0.4$~\cite{PhysRevD.88.084051,Musco_2005,escriva_primordial_2023,carr_constraints_2021},
the over-dense region collapse into a black hole.
The number density of black holes that collapse from Gaussian density perturbations
are determined by the Gaussian tail of the probability distribution function.
However, we emphasize that in our case, the amplitude of density perturbations induced by catalyzed PT has an upper limit,
which is determined by $p$ and $\gamma$.
If this upper limit is less than the threshold $\delta_c$,
no black holes are produced. Comparing to the result of \cite{jinno_super-slow_2023}, where they need $p\rightarrow 1$ and small $\gamma$ to produce PBHs, which is exactly confirmed by our result.


Following the Gaussian distribution of density perturbations, we get the variance of the density contrast
\begin{equation}
    \sigma_{\delta}^2(L)=\frac{p(1-p)}{(1+\frac{1}{1-\gamma}-p)^2\mathcal{H}^3(t_e)L^3}.
    \label{eqa:sigmaL}
\end{equation}
The variance is related to the power spectrum of curvature perturbations by~\cite{PhysRevD.81.104019}
\begin{equation}
    \sigma^2_{\delta}(L)=\frac{16}{81}\int d\ln{k}\,\mathcal{P}_{\mathcal{R}}(k)\exp{(-k^2L^2)}.
    \label{eqa:sigmaR}
\end{equation}
Here we have used the relation $\mathcal{P}_{\delta}(k)=\frac{16}{81}\mathcal{P}_{\mathcal{R}}(k)$ in comoving gauge at the horizon crossing time, which is our gauge choice as you can see in the Supplemental Material.
 It is proved that the power spectrum of density perturbations from the Poisson-type source scales as $k^3$ \cite{Zeng:2023jut}.
 We parameterize $\mathcal{P}_{\mathcal{R}}(k)$ in the form $\mathcal{P}_{\mathcal{R}}(k)=A_{b}(k/k_{cut})^{3}$
 where $k_{cut}$ is a cut-off scale arising from the requirement of the central limit theorem $L>S$.
 Since at smaller scales, the distribution of PBHs becomes non-Gaussian and $\mathcal{P}_{\mathcal{R}}(k)$ decreases rapidly,
 we simply adopt the approximation $k_{cut}=S^{-1}$ and $\mathcal{P}_{\mathcal{R}}(k)=0$ for $k>k_{cut}$.
 Then \eqref{eqa:sigmaR} is rewritten in the form
 \begin{eqnarray}
   \sigma_{\delta}^2(L) &=& \frac{16 A_{b}}{81 (k_{cut}L)^{3}}\int_{0}^{k_{cut}}d\left(kL\right)\exp{\left(-k^2 L^2 \right)}\left(kL\right)^{2} \nonumber \\
&=& \frac{4\sqrt{\pi} A_{b}}{81 (k_{cut}L)^{3}}\,,
\label{eqa:sigmaf}
 \end{eqnarray}
where
\begin{equation}
    A_b=\frac{81}{4\sqrt{\pi}}\frac{p^2(1-p)}{(1+\frac{1}{1-\gamma}-p)^2}.
    \label{eqa:A}
\end{equation}
The power spectrum of curvature perturbations induced by the catalyzed PT is given by
\begin{equation}
		\mathcal{P}_{\mathcal{R}}=\left\{
		\begin{aligned}
			&\frac{81}{4\sqrt{\pi}}\frac{p^2(1-p)}{(1+\frac{1}{1-\gamma}-p)^2}\left(\frac{k}{k_{cut}}\right)^3\quad &\mathrm{for}\; k\leq k_{cut}\,,\\
			&0\quad &\mathrm{for}\; k>k_{cut}\,,
		\end{aligned}
		\right.
\label{eqa:pr}
\end{equation}
 where $k_{cut}=n_{\mathrm{PBH}}^{\frac{1}{3}}(M)$ and $p$ is evaluated at the end of the PT,
 which is related to the comoving number density of seed PBHs and the energy scale of the first-order PT.
 The effects of $p$ and $\gamma$ on $\mathcal{P}_{\mathcal{R}}(k)$ are quite clear.
 If $\gamma$ approaches 1, which means that the catalyzing effect is negligible, then $\mathcal{P}_{\mathcal{R}}(k)$ is suppressed.
 When $p$ approaches $0$ or $1$, 
 the energy density differences are diluted at the scale $L$,
 so that the power spectrum of curvature perturbations is suppressed.
 Leaving these extreme case behind,
 we will see that there is a large parameter space
 in which a catalyzed PT may generate large curvature perturbations and even generate a detectable SGWB.
 The energy spectrum of the SGWB is roughly estimated by $\Omega_{GW}(k)\sim \mathcal{P}_{\mathcal{R}}^2(k)$.
 In this case the peak frequency of GWs is determined by the comoving number density of seed PBHs,
 and the amplitude is related to $A_b$ in~\eqref{eqa:A},
 which is both related to the abundance of seed PBHs and a specific PT model that characterized by $T_*$, $\alpha $ and $\beta/H_*$.

\emph{Specific example.}
 As a primary case of study, we focus on the first-order EWPT model.
 To provide an example, we need to first come up with a question:
 which kind of PBHs play a role as catalysts and generate curvature perturbations during the EWPT?
 To answer this question, we need two key quantities: the PBH masses and the PBH abundance.

\begin{figure}[h!]
    \includegraphics[width=3.5in]{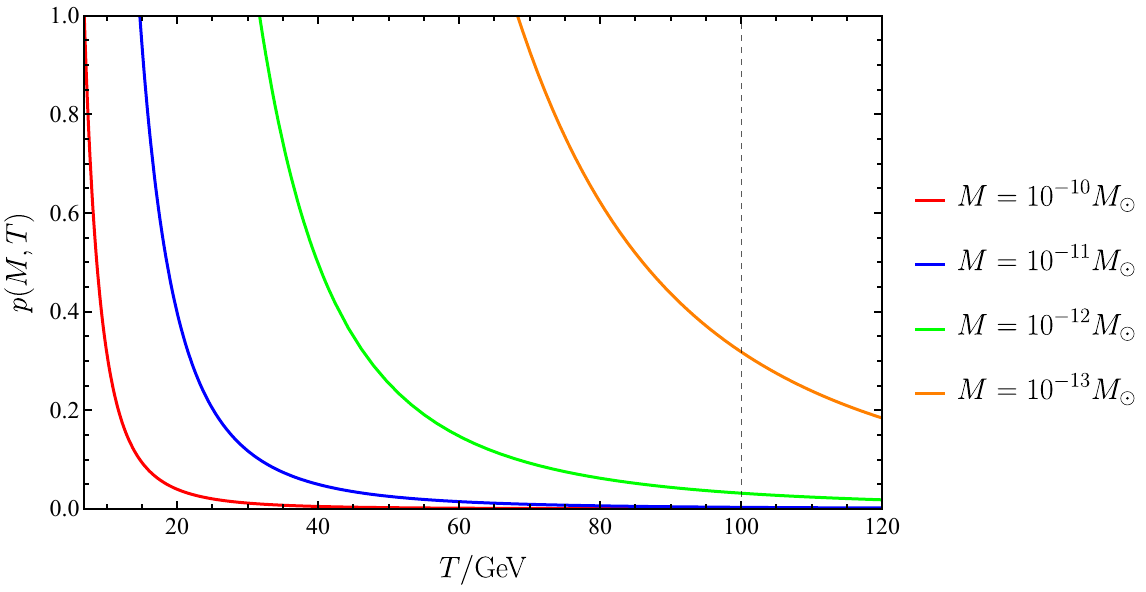}
    \caption{Average number of PBHs in a Hubble volume
    under the assumption that PBHs constitute all dark matter.
    The vertical dashed line corresponds to $T=100\,$GeV.}
    \label{fig:p}
\end{figure}

PBHs form in the early Universe.
It is known that PBHs form due to the collapse of the overdense region
during the radiation-dominated epoch and are the candidate of dark matter~\cite{hawking_black_1974,chapline_cosmological_1975,carr_constraints_2021,escriva_primordial_2023}.
Taking the assumption that PBHs are monochromatic,
it is reported that PBHs with masses around $10^{-15}M_{\odot}\sim 10^{-10}M_{\odot}$ can constitute all dark matter.
Because such a mass range of PBHs lies within the critical mass,
they work as catalysts during the EWPT. Their number density can be estimated by
\begin{equation}
  n_{\mathrm{PBH}}(M,t_0)\approx 3.285\times10^{10} \left(\frac{M_{\odot}}{M}\right)\rm{Mpc^{-1} },
  \label{eqa:num}
\end{equation}
where $t_0$ is the present time.
The average number of PBHs with masses $M$ in a Hubble-size region, $p(M,t)$,
is derived in~\eqref{eqa:p},
which is increasing over time due to the expansion of the Hubble horizon.
As mentioned above,
we are interested in the case of $p(M,T_{e})<1$,
where $T_e$ is the energy scale at the end of the EWPT.
We take PBHs with $M=10^{-13}M_{\odot}$ as an example,
the average number in a Hubble-size region is less than unity until the temperature decreases to $68\,$GeV.
We show PBHs that can be catalysts during the EWPT in Fig.~\ref{fig:p},
whose masses lie in the mass range of our interest.
If these PBHs form before the EWPT, the catalyzed EWPT is expected to generate large curvature perturbations at super-horizon scales.

\begin{figure}[h!]
    \includegraphics[width=3.5in]{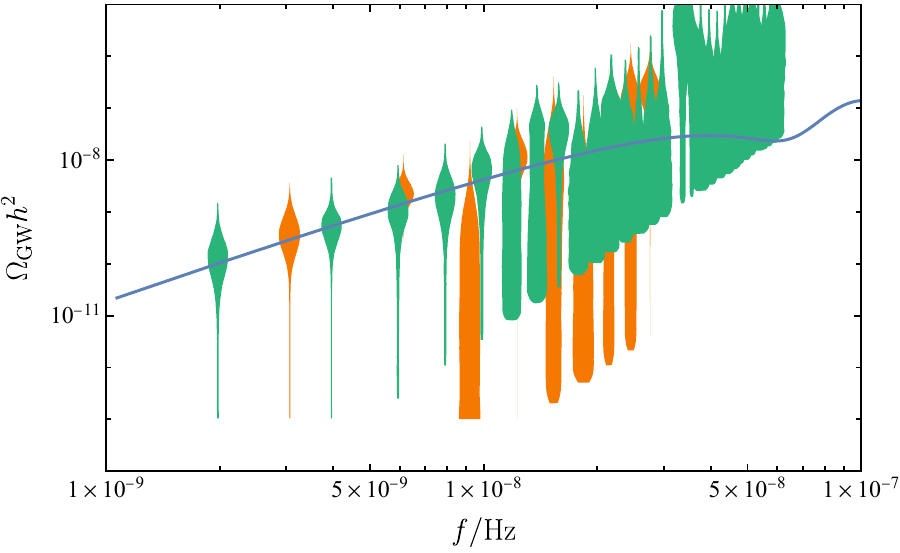}
    \caption{Energy spectrum of GWs (blue line) predicted by the catalyzed EWPT
    with the parameters $T_{e}=100\,\rm{GeV}$, $M=10^{-13}M_{\odot}$ and $\gamma=0.53$.
    The green and orange violins correspond to the SGWB observations from the NANOGrav and European pulsar timing array collaborations.}
    \label{fig:GW}
\end{figure}

It is an exciting coincidence 
that if the mass of PBHs is $M\sim 10^{-13}M_{\odot}$ which constitute all dark matter,
the SGWB from the EWPT catalyzed by PBHs
can explain the observed signal from the NANOGrav and European pulsar timing array.
By setting the energy scale at the end of the EWPT,
we calculate the average number of PBHs as catalysts in the Hubble volume by~\eqref{eqa:p}.
Fig.~\ref{fig:GW} shows the energy spectrum of GWs (blue line) predicted by the catalyzed EWPT
with $T_{e}=100\,\rm{GeV}$, $M=10^{-13}M_{\odot}$ and $\gamma=0.53$.
The green line in Fig.~\ref{fig:ga} corresponds to $\gamma=0.53$ in
the parameter space of $\alpha$ and $\beta/H_*$. 
Further more, we find that the upper limit of density perturbations are $\delta_{up}\sim0.176$,
which does not exceed the threshold of the PBH formation.
Therefore, curvature perturbations induce the SGWB to explain the pulsar timing array data~\cite{afzal_nanograv_2023,agazie_nanograv_2023,ellis_what_2023,gouttenoire_first-order_2023}
and naturally avoid the overproduction of PBHs.



\emph{Conclusions and discussions.}
We have investigated the first-order PT catalyzed by PBHs.
Due to the enhancement of the vacuum decay rate around a PBH,
the PT occurs earlier in the regions containing PBHs than in the regions without PBHs.
Such an asynchronism leads to super-horizon curvature perturbations
which induce the SGWB when re-entering the horizon in the radiation-dominated era.
If all dark matter is composed of PBHs with masses about $10^{-13}M_{\odot}$,
we find that the first-order EWPT catalyzed by PBHs
provides an explanation for the origin of the SGWB signal recently reported by pulsar timing array collaborations.
Usually PBHs are overproduced if the observed SGWB is induced by large curvature perturbations that follow the Gaussian distribution.
In our scenario, the upper limit of the amplitude of curvature perturbations is less than the threshold of the PBH formation.
Thus our scenario is free from the PBH overproduction problem.

PBHs in the mass range of $10^{-15}M_{\odot}\sim 10^{-10}M_{\odot}$ can account for all dark matter.
In our analysis we only consider PBHs with masses of $10^{-13}M_{\odot}$ as catalysts.
The increase of the PBH mass leads to a lower peak frequency of the energy spectrum of the SGWB
because the peak frequency of GWs is determined by the PBH number density
that is directly related to the PBH mass.
If PBHs constitute a part of dark matter, 
the fraction of dark matter in PBHs affects the peak frequency of GWs.
As the fraction decreases, the peak frequency becomes low for a fixed PBH mass.

In principle, a SGWB might be generated by bubble collisions, sound waves, and fluid turbulence during the PT.
The GW wavelengths of such a SGWB are shorter than the horizon at the PT time.
After the PT the SGWB is induced by curvature perturbations from the catalyzed PT. 
Therefore,
the peak frequency of the SGWB from the PT is higher than that of the SGWB from the catalyzed PT. 

In our work we consider PBHs as catalysts.
In principle, our formalism can be applied to other cases of impurity-catalyzed first-order PTs
as long as impurities are so sparse that their average number within a horizon volume is less than unity.

\emph{Acknowledgements}.
This work is supported in part by the National Key Research and Development Program
of China Grant No. 2020YFC2201501, in part by the National Natural Science Foundation
of China under Grant No. 12075297 and No. 12235019.

\bibliographystyle{apsrev4-2}
\bibliography{BHseedingPT}

\begin{thebibliography}{55}%
\makeatletter
\providecommand \@ifxundefined [1]{%
 \@ifx{#1\undefined}
}%
\providecommand \@ifnum [1]{%
 \ifnum #1\expandafter \@firstoftwo
 \else \expandafter \@secondoftwo
 \fi
}%
\providecommand \@ifx [1]{%
 \ifx #1\expandafter \@firstoftwo
 \else \expandafter \@secondoftwo
 \fi
}%
\providecommand \natexlab [1]{#1}%
\providecommand \enquote  [1]{``#1''}%
\providecommand \bibnamefont  [1]{#1}%
\providecommand \bibfnamefont [1]{#1}%
\providecommand \citenamefont [1]{#1}%
\providecommand \href@noop [0]{\@secondoftwo}%
\providecommand \href [0]{\begingroup \@sanitize@url \@href}%
\providecommand \@href[1]{\@@startlink{#1}\@@href}%
\providecommand \@@href[1]{\endgroup#1\@@endlink}%
\providecommand \@sanitize@url [0]{\catcode `\\12\catcode `\$12\catcode
  `\&12\catcode `\#12\catcode `\^12\catcode `\_12\catcode `\%12\relax}%
\providecommand \@@startlink[1]{}%
\providecommand \@@endlink[0]{}%
\providecommand \url  [0]{\begingroup\@sanitize@url \@url }%
\providecommand \@url [1]{\endgroup\@href {#1}{\urlprefix }}%
\providecommand \urlprefix  [0]{URL }%
\providecommand \Eprint [0]{\href }%
\providecommand \doibase [0]{https://doi.org/}%
\providecommand \selectlanguage [0]{\@gobble}%
\providecommand \bibinfo  [0]{\@secondoftwo}%
\providecommand \bibfield  [0]{\@secondoftwo}%
\providecommand \translation [1]{[#1]}%
\providecommand \BibitemOpen [0]{}%
\providecommand \bibitemStop [0]{}%
\providecommand \bibitemNoStop [0]{.\EOS\space}%
\providecommand \EOS [0]{\spacefactor3000\relax}%
\providecommand \BibitemShut  [1]{\csname bibitem#1\endcsname}%
\let\auto@bib@innerbib\@empty
\bibitem [{\citenamefont {Cline}\ and\ \citenamefont
  {Lemieux}(1997)}]{PhysRevD.55.3873}%
  \BibitemOpen
  \bibfield  {author} {\bibinfo {author} {\bibfnamefont {J.~M.}\ \bibnamefont
  {Cline}}\ and\ \bibinfo {author} {\bibfnamefont {P.-A.}\ \bibnamefont
  {Lemieux}},\ }\href {https://doi.org/10.1103/PhysRevD.55.3873} {\bibfield
  {journal} {\bibinfo  {journal} {Phys. Rev. D}\ }\textbf {\bibinfo {volume}
  {55}},\ \bibinfo {pages} {3873} (\bibinfo {year} {1997})}\BibitemShut
  {NoStop}%
\bibitem [{\citenamefont {Losada}(1997)}]{PhysRevD.56.2893}%
  \BibitemOpen
  \bibfield  {author} {\bibinfo {author} {\bibfnamefont {M.}~\bibnamefont
  {Losada}},\ }\href {https://doi.org/10.1103/PhysRevD.56.2893} {\bibfield
  {journal} {\bibinfo  {journal} {Phys. Rev. D}\ }\textbf {\bibinfo {volume}
  {56}},\ \bibinfo {pages} {2893} (\bibinfo {year} {1997})}\BibitemShut
  {NoStop}%
\bibitem [{\citenamefont {Bödeker}\ \emph {et~al.}(1997)\citenamefont
  {Bödeker}, \citenamefont {John}, \citenamefont {Laine},\ and\ \citenamefont
  {Schmidt}}]{BODEKER1997387}%
  \BibitemOpen
  \bibfield  {author} {\bibinfo {author} {\bibfnamefont {D.}~\bibnamefont
  {Bödeker}}, \bibinfo {author} {\bibfnamefont {P.}~\bibnamefont {John}},
  \bibinfo {author} {\bibfnamefont {M.}~\bibnamefont {Laine}},\ and\ \bibinfo
  {author} {\bibfnamefont {M.}~\bibnamefont {Schmidt}},\ }\href
  {https://doi.org/https://doi.org/10.1016/S0550-3213(97)00252-6} {\bibfield
  {journal} {\bibinfo  {journal} {Nuclear Physics B}\ }\textbf {\bibinfo
  {volume} {497}},\ \bibinfo {pages} {387} (\bibinfo {year}
  {1997})}\BibitemShut {NoStop}%
\bibitem [{\citenamefont {Laine}(1996)}]{LAINE199643}%
  \BibitemOpen
  \bibfield  {author} {\bibinfo {author} {\bibfnamefont {M.}~\bibnamefont
  {Laine}},\ }\href
  {https://doi.org/https://doi.org/10.1016/S0550-3213(96)90121-2} {\bibfield
  {journal} {\bibinfo  {journal} {Nuclear Physics B}\ }\textbf {\bibinfo
  {volume} {481}},\ \bibinfo {pages} {43} (\bibinfo {year} {1996})}\BibitemShut
  {NoStop}%
\bibitem [{\citenamefont {Kamionkowski}\ \emph {et~al.}(1994)\citenamefont
  {Kamionkowski}, \citenamefont {Kosowsky},\ and\ \citenamefont
  {Turner}}]{PhysRevD.49.2837}%
  \BibitemOpen
  \bibfield  {author} {\bibinfo {author} {\bibfnamefont {M.}~\bibnamefont
  {Kamionkowski}}, \bibinfo {author} {\bibfnamefont {A.}~\bibnamefont
  {Kosowsky}},\ and\ \bibinfo {author} {\bibfnamefont {M.~S.}\ \bibnamefont
  {Turner}},\ }\href {https://doi.org/10.1103/PhysRevD.49.2837} {\bibfield
  {journal} {\bibinfo  {journal} {Phys. Rev. D}\ }\textbf {\bibinfo {volume}
  {49}},\ \bibinfo {pages} {2837} (\bibinfo {year} {1994})}\BibitemShut
  {NoStop}%
\bibitem [{\citenamefont {Huber}\ and\ \citenamefont
  {Konstandin}(2008)}]{Huber_2008}%
  \BibitemOpen
  \bibfield  {author} {\bibinfo {author} {\bibfnamefont {S.~J.}\ \bibnamefont
  {Huber}}\ and\ \bibinfo {author} {\bibfnamefont {T.}~\bibnamefont
  {Konstandin}},\ }\href {https://doi.org/10.1088/1475-7516/2008/09/022}
  {\bibfield  {journal} {\bibinfo  {journal} {Journal of Cosmology and
  Astroparticle Physics}\ }\textbf {\bibinfo {volume} {2008}}\bibinfo  {number}
  { (09)},\ \bibinfo {pages} {022}}\BibitemShut {NoStop}%
\bibitem [{\citenamefont {Hindmarsh}\ \emph {et~al.}(2014)\citenamefont
  {Hindmarsh}, \citenamefont {Huber}, \citenamefont {Rummukainen},\ and\
  \citenamefont {Weir}}]{PhysRevLett.112.041301}%
  \BibitemOpen
\bibfield  {number} {  }\bibfield  {author} {\bibinfo {author} {\bibfnamefont
  {M.}~\bibnamefont {Hindmarsh}}, \bibinfo {author} {\bibfnamefont {S.~J.}\
  \bibnamefont {Huber}}, \bibinfo {author} {\bibfnamefont {K.}~\bibnamefont
  {Rummukainen}},\ and\ \bibinfo {author} {\bibfnamefont {D.~J.}\ \bibnamefont
  {Weir}},\ }\href {https://doi.org/10.1103/PhysRevLett.112.041301} {\bibfield
  {journal} {\bibinfo  {journal} {Phys. Rev. Lett.}\ }\textbf {\bibinfo
  {volume} {112}},\ \bibinfo {pages} {041301} (\bibinfo {year}
  {2014})}\BibitemShut {NoStop}%
\bibitem [{\citenamefont {Hindmarsh}\ \emph {et~al.}(2017)\citenamefont
  {Hindmarsh}, \citenamefont {Huber}, \citenamefont {Rummukainen},\ and\
  \citenamefont {Weir}}]{PhysRevD.96.103520}%
  \BibitemOpen
  \bibfield  {author} {\bibinfo {author} {\bibfnamefont {M.}~\bibnamefont
  {Hindmarsh}}, \bibinfo {author} {\bibfnamefont {S.~J.}\ \bibnamefont
  {Huber}}, \bibinfo {author} {\bibfnamefont {K.}~\bibnamefont {Rummukainen}},\
  and\ \bibinfo {author} {\bibfnamefont {D.~J.}\ \bibnamefont {Weir}},\ }\href
  {https://doi.org/10.1103/PhysRevD.96.103520} {\bibfield  {journal} {\bibinfo
  {journal} {Phys. Rev. D}\ }\textbf {\bibinfo {volume} {96}},\ \bibinfo
  {pages} {103520} (\bibinfo {year} {2017})}\BibitemShut {NoStop}%
\bibitem [{\citenamefont {Hindmarsh}\ \emph {et~al.}(2015)\citenamefont
  {Hindmarsh}, \citenamefont {Huber}, \citenamefont {Rummukainen},\ and\
  \citenamefont {Weir}}]{PhysRevD.92.123009}%
  \BibitemOpen
  \bibfield  {author} {\bibinfo {author} {\bibfnamefont {M.}~\bibnamefont
  {Hindmarsh}}, \bibinfo {author} {\bibfnamefont {S.~J.}\ \bibnamefont
  {Huber}}, \bibinfo {author} {\bibfnamefont {K.}~\bibnamefont {Rummukainen}},\
  and\ \bibinfo {author} {\bibfnamefont {D.~J.}\ \bibnamefont {Weir}},\ }\href
  {https://doi.org/10.1103/PhysRevD.92.123009} {\bibfield  {journal} {\bibinfo
  {journal} {Phys. Rev. D}\ }\textbf {\bibinfo {volume} {92}},\ \bibinfo
  {pages} {123009} (\bibinfo {year} {2015})}\BibitemShut {NoStop}%
\bibitem [{\citenamefont {Caprini}\ and\ \citenamefont
  {Durrer}(2006)}]{PhysRevD.74.063521}%
  \BibitemOpen
  \bibfield  {author} {\bibinfo {author} {\bibfnamefont {C.}~\bibnamefont
  {Caprini}}\ and\ \bibinfo {author} {\bibfnamefont {R.}~\bibnamefont
  {Durrer}},\ }\href {https://doi.org/10.1103/PhysRevD.74.063521} {\bibfield
  {journal} {\bibinfo  {journal} {Phys. Rev. D}\ }\textbf {\bibinfo {volume}
  {74}},\ \bibinfo {pages} {063521} (\bibinfo {year} {2006})}\BibitemShut
  {NoStop}%
\bibitem [{\citenamefont {Caprini}\ \emph {et~al.}(2009)\citenamefont
  {Caprini}, \citenamefont {Durrer},\ and\ \citenamefont
  {Servant}}]{ChiaraCaprini_2009}%
  \BibitemOpen
  \bibfield  {author} {\bibinfo {author} {\bibfnamefont {C.}~\bibnamefont
  {Caprini}}, \bibinfo {author} {\bibfnamefont {R.}~\bibnamefont {Durrer}},\
  and\ \bibinfo {author} {\bibfnamefont {G.}~\bibnamefont {Servant}},\ }\href
  {https://doi.org/10.1088/1475-7516/2009/12/024} {\bibfield  {journal}
  {\bibinfo  {journal} {Journal of Cosmology and Astroparticle Physics}\
  }\textbf {\bibinfo {volume} {2009}}\bibinfo  {number} { (12)},\ \bibinfo
  {pages} {024}}\BibitemShut {NoStop}%
\bibitem [{\citenamefont {Kosowsky}\ \emph {et~al.}(2002)\citenamefont
  {Kosowsky}, \citenamefont {Mack},\ and\ \citenamefont
  {Kahniashvili}}]{PhysRevD.66.024030}%
  \BibitemOpen
\bibfield  {number} {  }\bibfield  {author} {\bibinfo {author} {\bibfnamefont
  {A.}~\bibnamefont {Kosowsky}}, \bibinfo {author} {\bibfnamefont
  {A.}~\bibnamefont {Mack}},\ and\ \bibinfo {author} {\bibfnamefont
  {T.}~\bibnamefont {Kahniashvili}},\ }\href
  {https://doi.org/10.1103/PhysRevD.66.024030} {\bibfield  {journal} {\bibinfo
  {journal} {Phys. Rev. D}\ }\textbf {\bibinfo {volume} {66}},\ \bibinfo
  {pages} {024030} (\bibinfo {year} {2002})}\BibitemShut {NoStop}%
\bibitem [{\citenamefont {Gogoberidze}\ \emph {et~al.}(2007)\citenamefont
  {Gogoberidze}, \citenamefont {Kahniashvili},\ and\ \citenamefont
  {Kosowsky}}]{PhysRevD.76.083002}%
  \BibitemOpen
  \bibfield  {author} {\bibinfo {author} {\bibfnamefont {G.}~\bibnamefont
  {Gogoberidze}}, \bibinfo {author} {\bibfnamefont {T.}~\bibnamefont
  {Kahniashvili}},\ and\ \bibinfo {author} {\bibfnamefont {A.}~\bibnamefont
  {Kosowsky}},\ }\href {https://doi.org/10.1103/PhysRevD.76.083002} {\bibfield
  {journal} {\bibinfo  {journal} {Phys. Rev. D}\ }\textbf {\bibinfo {volume}
  {76}},\ \bibinfo {pages} {083002} (\bibinfo {year} {2007})}\BibitemShut
  {NoStop}%
\bibitem [{\citenamefont {Peter}\ \emph {et~al.}(2018)\citenamefont {Peter},
  \citenamefont {Martin},\ and\ \citenamefont {Günter}}]{Peter_2018}%
  \BibitemOpen
  \bibfield  {author} {\bibinfo {author} {\bibfnamefont {N.}~\bibnamefont
  {Peter}}, \bibinfo {author} {\bibfnamefont {S.}~\bibnamefont {Martin}},\ and\
  \bibinfo {author} {\bibfnamefont {S.}~\bibnamefont {Günter}},\ }\href
  {https://doi.org/10.1088/1361-6382/aac89c} {\bibfield  {journal} {\bibinfo
  {journal} {Classical and Quantum Gravity}\ }\textbf {\bibinfo {volume}
  {35}},\ \bibinfo {pages} {144001} (\bibinfo {year} {2018})}\BibitemShut
  {NoStop}%
\bibitem [{\citenamefont {Ellis}\ \emph {et~al.}(2023)\citenamefont {Ellis},
  \citenamefont {Fairbairn}, \citenamefont {Franciolini}, \citenamefont
  {Hütsi}, \citenamefont {Iovino~Jr.}, \citenamefont {Lewicki}, \citenamefont
  {Raidal}, \citenamefont {Urrutia}, \citenamefont {Vaskonen},\ and\
  \citenamefont {Veermäe}}]{ellis_what_2023}%
  \BibitemOpen
  \bibfield  {author} {\bibinfo {author} {\bibfnamefont {J.}~\bibnamefont
  {Ellis}}, \bibinfo {author} {\bibfnamefont {M.}~\bibnamefont {Fairbairn}},
  \bibinfo {author} {\bibfnamefont {G.}~\bibnamefont {Franciolini}}, \bibinfo
  {author} {\bibfnamefont {G.}~\bibnamefont {Hütsi}}, \bibinfo {author}
  {\bibfnamefont {A.}~\bibnamefont {Iovino~Jr.}}, \bibinfo {author}
  {\bibfnamefont {M.}~\bibnamefont {Lewicki}}, \bibinfo {author} {\bibfnamefont
  {M.}~\bibnamefont {Raidal}}, \bibinfo {author} {\bibfnamefont
  {J.}~\bibnamefont {Urrutia}}, \bibinfo {author} {\bibfnamefont
  {V.}~\bibnamefont {Vaskonen}},\ and\ \bibinfo {author} {\bibfnamefont
  {H.}~\bibnamefont {Veermäe}},\ }\href {http://arxiv.org/abs/2308.08546}
  {\bibinfo {title} {What is the source of the {PTA} {GW} signal?}} (\bibinfo
  {year} {2023}),\ \bibinfo {note} {arXiv:2308.08546 [astro-ph,
  physics:hep-th]}\BibitemShut {NoStop}%
\bibitem [{\citenamefont {et~al}(2023{\natexlab{a}})}]{afzal_nanograv_2023}%
  \BibitemOpen
  \bibfield  {author} {\bibinfo {author} {\bibfnamefont {A.}~\bibnamefont
  {et~al}},\ }\href@noop {} {\bibfield  {journal} {\bibinfo  {journal} {The
  Astrophysical Journal Letters}\ } (\bibinfo {year}
  {2023}{\natexlab{a}})}\BibitemShut {NoStop}%
\bibitem [{\citenamefont {et~al}(2023{\natexlab{b}})}]{agazie_nanograv_2023}%
  \BibitemOpen
  \bibfield  {author} {\bibinfo {author} {\bibfnamefont {A.}~\bibnamefont
  {et~al}},\ }\href {https://doi.org/10.3847/2041-8213/acdac6} {\bibinfo
  {title} {The {NANOGrav} 15-year {Data} {Set}: {Evidence} for a
  {Gravitational}-{Wave} {Background}}} (\bibinfo {year}
  {2023}{\natexlab{b}}),\ \bibinfo {note} {arXiv:2306.16213 [astro-ph,
  physics:gr-qc]}\BibitemShut {NoStop}%
\bibitem [{\citenamefont {Hiscock}(1987)}]{hiscock_can_1987}%
  \BibitemOpen
  \bibfield  {author} {\bibinfo {author} {\bibfnamefont {W.~A.}\ \bibnamefont
  {Hiscock}},\ }\href {https://doi.org/10.1103/PhysRevD.35.1161} {\bibfield
  {journal} {\bibinfo  {journal} {Physical Review D}\ }\textbf {\bibinfo
  {volume} {35}},\ \bibinfo {pages} {1161} (\bibinfo {year}
  {1987})}\BibitemShut {NoStop}%
\bibitem [{\citenamefont {Moss}(1985)}]{PhysRevD.32.1333}%
  \BibitemOpen
  \bibfield  {author} {\bibinfo {author} {\bibfnamefont {I.~G.}\ \bibnamefont
  {Moss}},\ }\href {https://doi.org/10.1103/PhysRevD.32.1333} {\bibfield
  {journal} {\bibinfo  {journal} {Phys. Rev. D}\ }\textbf {\bibinfo {volume}
  {32}},\ \bibinfo {pages} {1333} (\bibinfo {year} {1985})}\BibitemShut
  {NoStop}%
\bibitem [{\citenamefont {Gregory}\ \emph {et~al.}(2014)\citenamefont
  {Gregory}, \citenamefont {Moss},\ and\ \citenamefont
  {Withers}}]{gregory_black_2014}%
  \BibitemOpen
  \bibfield  {author} {\bibinfo {author} {\bibfnamefont {R.}~\bibnamefont
  {Gregory}}, \bibinfo {author} {\bibfnamefont {I.~G.}\ \bibnamefont {Moss}},\
  and\ \bibinfo {author} {\bibfnamefont {B.}~\bibnamefont {Withers}},\ }\href
  {https://doi.org/10.1007/JHEP03(2014)081} {\bibfield  {journal} {\bibinfo
  {journal} {Journal of High Energy Physics}\ }\textbf {\bibinfo {volume}
  {2014}},\ \bibinfo {pages} {81} (\bibinfo {year} {2014})},\ \bibinfo {note}
  {arXiv:1401.0017 [hep-th]}\BibitemShut {NoStop}%
\bibitem [{\citenamefont {Burda}\ \emph {et~al.}(2015)\citenamefont {Burda},
  \citenamefont {Gregory},\ and\ \citenamefont {Moss}}]{burda_vacuum_2015}%
  \BibitemOpen
  \bibfield  {author} {\bibinfo {author} {\bibfnamefont {P.}~\bibnamefont
  {Burda}}, \bibinfo {author} {\bibfnamefont {R.}~\bibnamefont {Gregory}},\
  and\ \bibinfo {author} {\bibfnamefont {I.}~\bibnamefont {Moss}},\ }\href
  {https://doi.org/10.1007/JHEP08(2015)114} {\bibfield  {journal} {\bibinfo
  {journal} {Journal of High Energy Physics}\ }\textbf {\bibinfo {volume}
  {2015}},\ \bibinfo {pages} {114} (\bibinfo {year} {2015})},\ \bibinfo {note}
  {arXiv:1503.07331 [gr-qc, physics:hep-ph, physics:hep-th]}\BibitemShut
  {NoStop}%
\bibitem [{\citenamefont {Burda}\ \emph {et~al.}(2016)\citenamefont {Burda},
  \citenamefont {Gregory},\ and\ \citenamefont {Moss}}]{burda_fate_2016}%
  \BibitemOpen
  \bibfield  {author} {\bibinfo {author} {\bibfnamefont {P.}~\bibnamefont
  {Burda}}, \bibinfo {author} {\bibfnamefont {R.}~\bibnamefont {Gregory}},\
  and\ \bibinfo {author} {\bibfnamefont {I.}~\bibnamefont {Moss}},\ }\href
  {https://doi.org/10.1007/JHEP06(2016)025} {\bibfield  {journal} {\bibinfo
  {journal} {Journal of High Energy Physics}\ }\textbf {\bibinfo {volume}
  {2016}},\ \bibinfo {pages} {25} (\bibinfo {year} {2016})},\ \bibinfo {note}
  {arXiv:1601.02152 [gr-qc, physics:hep-ph, physics:hep-th]}\BibitemShut
  {NoStop}%
\bibitem [{\citenamefont {Canko}\ \emph {et~al.}(2018)\citenamefont {Canko},
  \citenamefont {Gialamas}, \citenamefont {Jelic-Cizmek}, \citenamefont
  {Riotto},\ and\ \citenamefont {Tetradis}}]{Canko:2017ebb}%
  \BibitemOpen
  \bibfield  {author} {\bibinfo {author} {\bibfnamefont {D.}~\bibnamefont
  {Canko}}, \bibinfo {author} {\bibfnamefont {I.}~\bibnamefont {Gialamas}},
  \bibinfo {author} {\bibfnamefont {G.}~\bibnamefont {Jelic-Cizmek}}, \bibinfo
  {author} {\bibfnamefont {A.}~\bibnamefont {Riotto}},\ and\ \bibinfo {author}
  {\bibfnamefont {N.}~\bibnamefont {Tetradis}},\ }\href
  {https://doi.org/10.1140/epjc/s10052-018-5808-y} {\bibfield  {journal}
  {\bibinfo  {journal} {Eur. Phys. J. C}\ }\textbf {\bibinfo {volume} {78}},\
  \bibinfo {pages} {328} (\bibinfo {year} {2018})},\ \Eprint
  {https://arxiv.org/abs/1706.01364} {arXiv:1706.01364 [hep-th]} \BibitemShut
  {NoStop}%
\bibitem [{\citenamefont {Gregory}\ \emph
  {et~al.}(2020{\natexlab{a}})\citenamefont {Gregory}, \citenamefont {Moss},\
  and\ \citenamefont {Oshita}}]{gregory_black_2020}%
  \BibitemOpen
  \bibfield  {author} {\bibinfo {author} {\bibfnamefont {R.}~\bibnamefont
  {Gregory}}, \bibinfo {author} {\bibfnamefont {I.~G.}\ \bibnamefont {Moss}},\
  and\ \bibinfo {author} {\bibfnamefont {N.}~\bibnamefont {Oshita}},\ }\href
  {https://doi.org/10.1007/JHEP07(2020)024} {\bibfield  {journal} {\bibinfo
  {journal} {Journal of High Energy Physics}\ }\textbf {\bibinfo {volume}
  {2020}},\ \bibinfo {pages} {24} (\bibinfo {year} {2020}{\natexlab{a}})},\
  \bibinfo {note} {arXiv:2003.04927 [gr-qc, physics:hep-th]}\BibitemShut
  {NoStop}%
\bibitem [{\citenamefont {Dai}\ \emph {et~al.}(2020)\citenamefont {Dai},
  \citenamefont {Gregory},\ and\ \citenamefont
  {Stojkovic}}]{PhysRevD.101.125012}%
  \BibitemOpen
  \bibfield  {author} {\bibinfo {author} {\bibfnamefont {D.-C.}\ \bibnamefont
  {Dai}}, \bibinfo {author} {\bibfnamefont {R.}~\bibnamefont {Gregory}},\ and\
  \bibinfo {author} {\bibfnamefont {D.}~\bibnamefont {Stojkovic}},\ }\href
  {https://doi.org/10.1103/PhysRevD.101.125012} {\bibfield  {journal} {\bibinfo
   {journal} {Phys. Rev. D}\ }\textbf {\bibinfo {volume} {101}},\ \bibinfo
  {pages} {125012} (\bibinfo {year} {2020})}\BibitemShut {NoStop}%
\bibitem [{\citenamefont {Gregory}\ \emph
  {et~al.}(2020{\natexlab{b}})\citenamefont {Gregory}, \citenamefont {Moss},
  \citenamefont {Oshita},\ and\ \citenamefont
  {Patrick}}]{gregory_hawking-moss_2020}%
  \BibitemOpen
  \bibfield  {author} {\bibinfo {author} {\bibfnamefont {R.}~\bibnamefont
  {Gregory}}, \bibinfo {author} {\bibfnamefont {I.~G.}\ \bibnamefont {Moss}},
  \bibinfo {author} {\bibfnamefont {N.}~\bibnamefont {Oshita}},\ and\ \bibinfo
  {author} {\bibfnamefont {S.}~\bibnamefont {Patrick}},\ }\href
  {https://doi.org/10.1007/JHEP09(2020)135} {\bibfield  {journal} {\bibinfo
  {journal} {Journal of High Energy Physics}\ }\textbf {\bibinfo {volume}
  {2020}},\ \bibinfo {pages} {135} (\bibinfo {year} {2020}{\natexlab{b}})},\
  \bibinfo {note} {arXiv:2007.11428 [gr-qc, physics:hep-th]}\BibitemShut
  {NoStop}%
\bibitem [{\citenamefont {Jinno}\ \emph {et~al.}(2023)\citenamefont {Jinno},
  \citenamefont {Kume},\ and\ \citenamefont {Yamada}}]{jinno_super-slow_2023}%
  \BibitemOpen
  \bibfield  {author} {\bibinfo {author} {\bibfnamefont {R.}~\bibnamefont
  {Jinno}}, \bibinfo {author} {\bibfnamefont {J.}~\bibnamefont {Kume}},\ and\
  \bibinfo {author} {\bibfnamefont {M.}~\bibnamefont {Yamada}},\ }\href
  {http://arxiv.org/abs/2310.06901} {\bibinfo {title} {Super-slow phase
  transition catalyzed by {BHs} and the birth of baby {BHs}}} (\bibinfo {year}
  {2023}),\ \bibinfo {note} {arXiv:2310.06901 [astro-ph, physics:gr-qc,
  physics:hep-ph]}\BibitemShut {NoStop}%
\bibitem [{\citenamefont {Mukaida}\ and\ \citenamefont
  {Yamada}(2017)}]{mukaida_false_2017}%
  \BibitemOpen
  \bibfield  {author} {\bibinfo {author} {\bibfnamefont {K.}~\bibnamefont
  {Mukaida}}\ and\ \bibinfo {author} {\bibfnamefont {M.}~\bibnamefont
  {Yamada}},\ }\href {https://doi.org/10.1103/PhysRevD.96.103514} {\bibfield
  {journal} {\bibinfo  {journal} {Physical Review D}\ }\textbf {\bibinfo
  {volume} {96}},\ \bibinfo {pages} {103514} (\bibinfo {year} {2017})},\
  \bibinfo {note} {arXiv:1706.04523 [gr-qc, physics:hep-ph,
  physics:hep-th]}\BibitemShut {NoStop}%
\bibitem [{\citenamefont {Kohri}\ and\ \citenamefont
  {Matsui}(2018)}]{kohri_electroweak_2018}%
  \BibitemOpen
  \bibfield  {author} {\bibinfo {author} {\bibfnamefont {K.}~\bibnamefont
  {Kohri}}\ and\ \bibinfo {author} {\bibfnamefont {H.}~\bibnamefont {Matsui}},\
  }\href {https://doi.org/10.1103/PhysRevD.98.123509} {\bibfield  {journal}
  {\bibinfo  {journal} {Physical Review D}\ }\textbf {\bibinfo {volume} {98}},\
  \bibinfo {pages} {123509} (\bibinfo {year} {2018})},\ \bibinfo {note}
  {arXiv:1708.02138 [astro-ph, physics:gr-qc, physics:hep-ph,
  physics:hep-th]}\BibitemShut {NoStop}%
\bibitem [{\citenamefont {El-Menoufi}\ \emph {et~al.}(2020)\citenamefont
  {El-Menoufi}, \citenamefont {Huber},\ and\ \citenamefont
  {Manuel}}]{el-menoufi_black_2020}%
  \BibitemOpen
  \bibfield  {author} {\bibinfo {author} {\bibfnamefont {B.~K.}\ \bibnamefont
  {El-Menoufi}}, \bibinfo {author} {\bibfnamefont {S.~J.}\ \bibnamefont
  {Huber}},\ and\ \bibinfo {author} {\bibfnamefont {J.~P.}\ \bibnamefont
  {Manuel}},\ }\href {http://arxiv.org/abs/2006.16275} {\bibinfo {title} {Black
  holes seeding cosmological phase transitions}} (\bibinfo {year} {2020}),\
  \bibinfo {note} {arXiv:2006.16275 [gr-qc, physics:hep-ph,
  physics:hep-th]}\BibitemShut {NoStop}%
\bibitem [{\citenamefont {Hayashi}\ \emph {et~al.}(2020)\citenamefont
  {Hayashi}, \citenamefont {Kamada}, \citenamefont {Oshita},\ and\
  \citenamefont {Yokoyama}}]{hayashi_catalyzed_2020}%
  \BibitemOpen
  \bibfield  {author} {\bibinfo {author} {\bibfnamefont {T.}~\bibnamefont
  {Hayashi}}, \bibinfo {author} {\bibfnamefont {K.}~\bibnamefont {Kamada}},
  \bibinfo {author} {\bibfnamefont {N.}~\bibnamefont {Oshita}},\ and\ \bibinfo
  {author} {\bibfnamefont {J.}~\bibnamefont {Yokoyama}},\ }\href
  {https://doi.org/10.1007/JHEP08(2020)088} {\bibfield  {journal} {\bibinfo
  {journal} {Journal of High Energy Physics}\ }\textbf {\bibinfo {volume}
  {2020}},\ \bibinfo {pages} {88} (\bibinfo {year} {2020})},\ \bibinfo {note}
  {arXiv:2005.12808 [astro-ph, physics:gr-qc, physics:hep-ph,
  physics:hep-th]}\BibitemShut {NoStop}%
\bibitem [{\citenamefont {Hamaide}\ \emph {et~al.}(2023)\citenamefont
  {Hamaide}, \citenamefont {Heurtier}, \citenamefont {Hu},\ and\ \citenamefont
  {Cheek}}]{hamaide_primordial_2023}%
  \BibitemOpen
  \bibfield  {author} {\bibinfo {author} {\bibfnamefont {L.}~\bibnamefont
  {Hamaide}}, \bibinfo {author} {\bibfnamefont {L.}~\bibnamefont {Heurtier}},
  \bibinfo {author} {\bibfnamefont {S.-Q.}\ \bibnamefont {Hu}},\ and\ \bibinfo
  {author} {\bibfnamefont {A.}~\bibnamefont {Cheek}},\ }\href
  {http://arxiv.org/abs/2311.01869} {\bibinfo {title} {Primordial {Black}
  {Holes} {Are} {True} {Vacuum} {Nurseries}}} (\bibinfo {year} {2023}),\
  \bibinfo {note} {arXiv:2311.01869 [astro-ph, physics:gr-qc, physics:hep-ph,
  physics:hep-th]}\BibitemShut {NoStop}%
\bibitem [{\citenamefont {Strumia}(2023)}]{strumia_black_2023}%
  \BibitemOpen
  \bibfield  {author} {\bibinfo {author} {\bibfnamefont {A.}~\bibnamefont
  {Strumia}},\ }\href {https://doi.org/10.1007/JHEP03(2023)039} {\bibfield
  {journal} {\bibinfo  {journal} {Journal of High Energy Physics}\ }\textbf
  {\bibinfo {volume} {2023}},\ \bibinfo {pages} {39} (\bibinfo {year}
  {2023})},\ \bibinfo {note} {arXiv:2209.05504 [hep-ph,
  physics:hep-th]}\BibitemShut {NoStop}%
\bibitem [{\citenamefont {Oshita}\ \emph {et~al.}(2019)\citenamefont {Oshita},
  \citenamefont {Yamada},\ and\ \citenamefont {Yamaguchi}}]{OSHITA2019149}%
  \BibitemOpen
  \bibfield  {author} {\bibinfo {author} {\bibfnamefont {N.}~\bibnamefont
  {Oshita}}, \bibinfo {author} {\bibfnamefont {M.}~\bibnamefont {Yamada}},\
  and\ \bibinfo {author} {\bibfnamefont {M.}~\bibnamefont {Yamaguchi}},\ }\href
  {https://doi.org/https://doi.org/10.1016/j.physletb.2019.02.032} {\bibfield
  {journal} {\bibinfo  {journal} {Physics Letters B}\ }\textbf {\bibinfo
  {volume} {791}},\ \bibinfo {pages} {149} (\bibinfo {year}
  {2019})}\BibitemShut {NoStop}%
\bibitem [{\citenamefont {Dine}\ \emph {et~al.}(1992)\citenamefont {Dine},
  \citenamefont {Leigh}, \citenamefont {Huet}, \citenamefont {Linde},\ and\
  \citenamefont {Linde}}]{dine_towards_1992}%
  \BibitemOpen
  \bibfield  {author} {\bibinfo {author} {\bibfnamefont {M.}~\bibnamefont
  {Dine}}, \bibinfo {author} {\bibfnamefont {R.}~\bibnamefont {Leigh}},
  \bibinfo {author} {\bibfnamefont {P.}~\bibnamefont {Huet}}, \bibinfo {author}
  {\bibfnamefont {A.}~\bibnamefont {Linde}},\ and\ \bibinfo {author}
  {\bibfnamefont {D.}~\bibnamefont {Linde}},\ }\href
  {https://doi.org/10.1103/PhysRevD.46.550} {\bibfield  {journal} {\bibinfo
  {journal} {Physical Review D}\ }\textbf {\bibinfo {volume} {46}},\ \bibinfo
  {pages} {550} (\bibinfo {year} {1992})},\ \bibinfo {note}
  {arXiv:hep-ph/9203203}\BibitemShut {NoStop}%
\bibitem [{\citenamefont {Ellis}\ \emph {et~al.}(2019)\citenamefont {Ellis},
  \citenamefont {Lewicki},\ and\ \citenamefont {No}}]{ellis_maximal_2019}%
  \BibitemOpen
  \bibfield  {author} {\bibinfo {author} {\bibfnamefont {J.}~\bibnamefont
  {Ellis}}, \bibinfo {author} {\bibfnamefont {M.}~\bibnamefont {Lewicki}},\
  and\ \bibinfo {author} {\bibfnamefont {J.~M.}\ \bibnamefont {No}},\ }\href
  {https://doi.org/10.1088/1475-7516/2019/04/003} {\bibfield  {journal}
  {\bibinfo  {journal} {Journal of Cosmology and Astroparticle Physics}\
  }\textbf {\bibinfo {volume} {2019}}\bibfield  {number} {\bibinfo  {number} {
  (04)},\ \bibinfo {pages} {003}},\ }\bibinfo {note} {arXiv:1809.08242
  [astro-ph, physics:hep-ph]}\BibitemShut {NoStop}%
\bibitem [{\citenamefont {Cai}\ \emph {et~al.}(2017)\citenamefont {Cai},
  \citenamefont {Cao}, \citenamefont {Guo}, \citenamefont {Wang},\ and\
  \citenamefont {Yang}}]{cai_gravitational-wave_2017}%
  \BibitemOpen
  \bibfield  {author} {\bibinfo {author} {\bibfnamefont {R.-G.}\ \bibnamefont
  {Cai}}, \bibinfo {author} {\bibfnamefont {Z.}~\bibnamefont {Cao}}, \bibinfo
  {author} {\bibfnamefont {Z.-K.}\ \bibnamefont {Guo}}, \bibinfo {author}
  {\bibfnamefont {S.-J.}\ \bibnamefont {Wang}},\ and\ \bibinfo {author}
  {\bibfnamefont {T.}~\bibnamefont {Yang}},\ }\href
  {https://doi.org/10.1093/nsr/nwx029} {\bibfield  {journal} {\bibinfo
  {journal} {National Science Review}\ }\textbf {\bibinfo {volume} {4}},\
  \bibinfo {pages} {687} (\bibinfo {year} {2017})},\ \bibinfo {note} {\_eprint:
  https://academic.oup.com/nsr/article-pdf/4/5/687/31566614/nwx029.pdf}\BibitemShut
  {NoStop}%
\bibitem [{\citenamefont {Carr}\ \emph {et~al.}(2021)\citenamefont {Carr},
  \citenamefont {Kohri}, \citenamefont {Sendouda},\ and\ \citenamefont
  {Yokoyama}}]{carr_constraints_2021}%
  \BibitemOpen
  \bibfield  {author} {\bibinfo {author} {\bibfnamefont {B.}~\bibnamefont
  {Carr}}, \bibinfo {author} {\bibfnamefont {K.}~\bibnamefont {Kohri}},
  \bibinfo {author} {\bibfnamefont {Y.}~\bibnamefont {Sendouda}},\ and\
  \bibinfo {author} {\bibfnamefont {J.}~\bibnamefont {Yokoyama}},\ }\href
  {https://doi.org/10.1088/1361-6633/ac1e31} {\bibfield  {journal} {\bibinfo
  {journal} {Reports on Progress in Physics}\ }\textbf {\bibinfo {volume}
  {84}},\ \bibinfo {pages} {116902} (\bibinfo {year} {2021})},\ \bibinfo {note}
  {arXiv:2002.12778 [astro-ph, physics:gr-qc, physics:hep-ph,
  physics:hep-th]}\BibitemShut {NoStop}%
\bibitem [{\citenamefont {Coleman}(1977)}]{PhysRevD.16.1248}%
  \BibitemOpen
  \bibfield  {author} {\bibinfo {author} {\bibfnamefont {S.}~\bibnamefont
  {Coleman}},\ }\href {https://doi.org/10.1103/PhysRevD.16.1248} {\bibfield
  {journal} {\bibinfo  {journal} {Phys. Rev. D}\ }\textbf {\bibinfo {volume}
  {16}},\ \bibinfo {pages} {1248} (\bibinfo {year} {1977})}\BibitemShut
  {NoStop}%
\bibitem [{\citenamefont {Enqvist}\ \emph {et~al.}(1992)\citenamefont
  {Enqvist}, \citenamefont {Ignatius}, \citenamefont {Kajantie},\ and\
  \citenamefont {Rummukainen}}]{PhysRevD.45.3415}%
  \BibitemOpen
  \bibfield  {author} {\bibinfo {author} {\bibfnamefont {K.}~\bibnamefont
  {Enqvist}}, \bibinfo {author} {\bibfnamefont {J.}~\bibnamefont {Ignatius}},
  \bibinfo {author} {\bibfnamefont {K.}~\bibnamefont {Kajantie}},\ and\
  \bibinfo {author} {\bibfnamefont {K.}~\bibnamefont {Rummukainen}},\ }\href
  {https://doi.org/10.1103/PhysRevD.45.3415} {\bibfield  {journal} {\bibinfo
  {journal} {Phys. Rev. D}\ }\textbf {\bibinfo {volume} {45}},\ \bibinfo
  {pages} {3415} (\bibinfo {year} {1992})}\BibitemShut {NoStop}%
\bibitem [{\citenamefont {Turner}\ \emph {et~al.}(1992)\citenamefont {Turner},
  \citenamefont {Weinberg},\ and\ \citenamefont {Widrow}}]{PhysRevD.46.2384}%
  \BibitemOpen
  \bibfield  {author} {\bibinfo {author} {\bibfnamefont {M.~S.}\ \bibnamefont
  {Turner}}, \bibinfo {author} {\bibfnamefont {E.~J.}\ \bibnamefont
  {Weinberg}},\ and\ \bibinfo {author} {\bibfnamefont {L.~M.}\ \bibnamefont
  {Widrow}},\ }\href {https://doi.org/10.1103/PhysRevD.46.2384} {\bibfield
  {journal} {\bibinfo  {journal} {Phys. Rev. D}\ }\textbf {\bibinfo {volume}
  {46}},\ \bibinfo {pages} {2384} (\bibinfo {year} {1992})}\BibitemShut
  {NoStop}%
\bibitem [{\citenamefont {Lyth}\ \emph {et~al.}(2005)\citenamefont {Lyth},
  \citenamefont {Malik},\ and\ \citenamefont {Sasaki}}]{Lyth:2004gb}%
  \BibitemOpen
  \bibfield  {author} {\bibinfo {author} {\bibfnamefont {D.~H.}\ \bibnamefont
  {Lyth}}, \bibinfo {author} {\bibfnamefont {K.~A.}\ \bibnamefont {Malik}},\
  and\ \bibinfo {author} {\bibfnamefont {M.}~\bibnamefont {Sasaki}},\ }\href
  {https://doi.org/10.1088/1475-7516/2005/05/004} {\bibfield  {journal}
  {\bibinfo  {journal} {JCAP}\ }\textbf {\bibinfo {volume} {05}},\ \bibinfo
  {pages} {004}},\ \Eprint {https://arxiv.org/abs/astro-ph/0411220}
  {arXiv:astro-ph/0411220} \BibitemShut {NoStop}%
\bibitem [{\citenamefont {Liu}\ \emph {et~al.}(2022)\citenamefont {Liu},
  \citenamefont {Bian}, \citenamefont {Cai}, \citenamefont {Guo},\ and\
  \citenamefont {Wang}}]{liu_primordial_2022}%
  \BibitemOpen
  \bibfield  {author} {\bibinfo {author} {\bibfnamefont {J.}~\bibnamefont
  {Liu}}, \bibinfo {author} {\bibfnamefont {L.}~\bibnamefont {Bian}}, \bibinfo
  {author} {\bibfnamefont {R.-G.}\ \bibnamefont {Cai}}, \bibinfo {author}
  {\bibfnamefont {Z.-K.}\ \bibnamefont {Guo}},\ and\ \bibinfo {author}
  {\bibfnamefont {S.-J.}\ \bibnamefont {Wang}},\ }\href
  {https://doi.org/10.1103/PhysRevD.105.L021303} {\bibfield  {journal}
  {\bibinfo  {journal} {Physical Review D}\ }\textbf {\bibinfo {volume}
  {105}},\ \bibinfo {pages} {L021303} (\bibinfo {year} {2022})},\ \bibinfo
  {note} {arXiv:2106.05637 [astro-ph, physics:gr-qc,
  physics:hep-ph]}\BibitemShut {NoStop}%
\bibitem [{\citenamefont {Liu}\ \emph {et~al.}(2023)\citenamefont {Liu},
  \citenamefont {Bian}, \citenamefont {Cai}, \citenamefont {Guo},\ and\
  \citenamefont {Wang}}]{liu_constraining_2023}%
  \BibitemOpen
  \bibfield  {author} {\bibinfo {author} {\bibfnamefont {J.}~\bibnamefont
  {Liu}}, \bibinfo {author} {\bibfnamefont {L.}~\bibnamefont {Bian}}, \bibinfo
  {author} {\bibfnamefont {R.-G.}\ \bibnamefont {Cai}}, \bibinfo {author}
  {\bibfnamefont {Z.-K.}\ \bibnamefont {Guo}},\ and\ \bibinfo {author}
  {\bibfnamefont {S.-J.}\ \bibnamefont {Wang}},\ }\href
  {https://doi.org/10.1103/PhysRevLett.130.051001} {\bibfield  {journal}
  {\bibinfo  {journal} {Physical Review Letters}\ }\textbf {\bibinfo {volume}
  {130}},\ \bibinfo {pages} {051001} (\bibinfo {year} {2023})},\ \bibinfo
  {note} {arXiv:2208.14086 [astro-ph, physics:gr-qc,
  physics:hep-ph]}\BibitemShut {NoStop}%
\bibitem [{\citenamefont {Harada}\ \emph {et~al.}(2015)\citenamefont {Harada},
  \citenamefont {Yoo}, \citenamefont {Nakama},\ and\ \citenamefont
  {Koga}}]{PhysRevD.91.084057}%
  \BibitemOpen
  \bibfield  {author} {\bibinfo {author} {\bibfnamefont {T.}~\bibnamefont
  {Harada}}, \bibinfo {author} {\bibfnamefont {C.-M.}\ \bibnamefont {Yoo}},
  \bibinfo {author} {\bibfnamefont {T.}~\bibnamefont {Nakama}},\ and\ \bibinfo
  {author} {\bibfnamefont {Y.}~\bibnamefont {Koga}},\ }\href
  {https://doi.org/10.1103/PhysRevD.91.084057} {\bibfield  {journal} {\bibinfo
  {journal} {Phys. Rev. D}\ }\textbf {\bibinfo {volume} {91}},\ \bibinfo
  {pages} {084057} (\bibinfo {year} {2015})}\BibitemShut {NoStop}%
\bibitem [{\citenamefont {Escrivà}\ \emph {et~al.}(2023)\citenamefont
  {Escrivà}, \citenamefont {Kuhnel},\ and\ \citenamefont
  {Tada}}]{escriva_primordial_2023}%
  \BibitemOpen
  \bibfield  {author} {\bibinfo {author} {\bibfnamefont {A.}~\bibnamefont
  {Escrivà}}, \bibinfo {author} {\bibfnamefont {F.}~\bibnamefont {Kuhnel}},\
  and\ \bibinfo {author} {\bibfnamefont {Y.}~\bibnamefont {Tada}},\ }\href
  {http://arxiv.org/abs/2211.05767} {\bibinfo {title} {Primordial {Black}
  {Holes}}} (\bibinfo {year} {2023}),\ \bibinfo {note} {arXiv:2211.05767
  [astro-ph, physics:gr-qc, physics:hep-ph, physics:hep-th]}\BibitemShut
  {NoStop}%
\bibitem [{\citenamefont {A.~Vilenkin}(1994)}]{vilenkin}%
  \BibitemOpen
  \bibfield  {author} {\bibinfo {author} {\bibfnamefont {E.~P. S.~S.}\
  \bibnamefont {A.~Vilenkin}, \bibfnamefont {Alexander~Vilenkin}},\ }\bibinfo
  {title} {Cosmic strings and other topological defects}\ (\bibinfo
  {publisher} {Cambridge University Press},\ \bibinfo {year} {1994})\ Chap.\
  \bibinfo {chapter} {13.3.2 A wall-dominated universe}\BibitemShut {NoStop}%
\bibitem [{\citenamefont {Gouttenoire}\ and\ \citenamefont
  {Volansky}(2023)}]{gouttenoire2023primordial}%
  \BibitemOpen
  \bibfield  {author} {\bibinfo {author} {\bibfnamefont {Y.}~\bibnamefont
  {Gouttenoire}}\ and\ \bibinfo {author} {\bibfnamefont {T.}~\bibnamefont
  {Volansky}},\ }\href@noop {} {\bibinfo {title} {Primordial black holes from
  supercooled phase transitions}} (\bibinfo {year} {2023}),\ \Eprint
  {https://arxiv.org/abs/2305.04942} {arXiv:2305.04942 [hep-ph]} \BibitemShut
  {NoStop}%
\bibitem [{\citenamefont {Harada}\ \emph {et~al.}(2013)\citenamefont {Harada},
  \citenamefont {Yoo},\ and\ \citenamefont {Kohri}}]{PhysRevD.88.084051}%
  \BibitemOpen
  \bibfield  {author} {\bibinfo {author} {\bibfnamefont {T.}~\bibnamefont
  {Harada}}, \bibinfo {author} {\bibfnamefont {C.-M.}\ \bibnamefont {Yoo}},\
  and\ \bibinfo {author} {\bibfnamefont {K.}~\bibnamefont {Kohri}},\ }\href
  {https://doi.org/10.1103/PhysRevD.88.084051} {\bibfield  {journal} {\bibinfo
  {journal} {Phys. Rev. D}\ }\textbf {\bibinfo {volume} {88}},\ \bibinfo
  {pages} {084051} (\bibinfo {year} {2013})}\BibitemShut {NoStop}%
\bibitem [{\citenamefont {Musco}\ \emph {et~al.}(2005)\citenamefont {Musco},
  \citenamefont {Miller},\ and\ \citenamefont {Rezzolla}}]{Musco_2005}%
  \BibitemOpen
  \bibfield  {author} {\bibinfo {author} {\bibfnamefont {I.}~\bibnamefont
  {Musco}}, \bibinfo {author} {\bibfnamefont {J.~C.}\ \bibnamefont {Miller}},\
  and\ \bibinfo {author} {\bibfnamefont {L.}~\bibnamefont {Rezzolla}},\ }\href
  {https://doi.org/10.1088/0264-9381/22/7/013} {\bibfield  {journal} {\bibinfo
  {journal} {Classical and Quantum Gravity}\ }\textbf {\bibinfo {volume}
  {22}},\ \bibinfo {pages} {1405} (\bibinfo {year} {2005})}\BibitemShut
  {NoStop}%
\bibitem [{\citenamefont {Carr}\ \emph {et~al.}(2010)\citenamefont {Carr},
  \citenamefont {Kohri}, \citenamefont {Sendouda},\ and\ \citenamefont
  {Yokoyama}}]{PhysRevD.81.104019}%
  \BibitemOpen
  \bibfield  {author} {\bibinfo {author} {\bibfnamefont {B.~J.}\ \bibnamefont
  {Carr}}, \bibinfo {author} {\bibfnamefont {K.}~\bibnamefont {Kohri}},
  \bibinfo {author} {\bibfnamefont {Y.}~\bibnamefont {Sendouda}},\ and\
  \bibinfo {author} {\bibfnamefont {J.}~\bibnamefont {Yokoyama}},\ }\href
  {https://doi.org/10.1103/PhysRevD.81.104019} {\bibfield  {journal} {\bibinfo
  {journal} {Phys. Rev. D}\ }\textbf {\bibinfo {volume} {81}},\ \bibinfo
  {pages} {104019} (\bibinfo {year} {2010})}\BibitemShut {NoStop}%
\bibitem [{\citenamefont {Zeng}\ \emph {et~al.}(2023)\citenamefont {Zeng},
  \citenamefont {Liu},\ and\ \citenamefont {Guo}}]{Zeng:2023jut}%
  \BibitemOpen
  \bibfield  {author} {\bibinfo {author} {\bibfnamefont {Z.-M.}\ \bibnamefont
  {Zeng}}, \bibinfo {author} {\bibfnamefont {J.}~\bibnamefont {Liu}},\ and\
  \bibinfo {author} {\bibfnamefont {Z.-K.}\ \bibnamefont {Guo}},\ }\href
  {https://doi.org/10.1103/PhysRevD.108.063005} {\bibfield  {journal} {\bibinfo
   {journal} {Phys. Rev. D}\ }\textbf {\bibinfo {volume} {108}},\ \bibinfo
  {pages} {063005} (\bibinfo {year} {2023})},\ \Eprint
  {https://arxiv.org/abs/2301.07230} {arXiv:2301.07230 [astro-ph.CO]}
  \BibitemShut {NoStop}%
\bibitem [{\citenamefont {HAWKING}(1974)}]{hawking_black_1974}%
  \BibitemOpen
  \bibfield  {author} {\bibinfo {author} {\bibfnamefont {S.~W.}\ \bibnamefont
  {HAWKING}},\ }\href {https://doi.org/10.1038/248030a0} {\bibfield  {journal}
  {\bibinfo  {journal} {Nature}\ }\textbf {\bibinfo {volume} {248}},\ \bibinfo
  {pages} {30} (\bibinfo {year} {1974})}\BibitemShut {NoStop}%
\bibitem [{\citenamefont {CHAPLINE}(1975)}]{chapline_cosmological_1975}%
  \BibitemOpen
  \bibfield  {author} {\bibinfo {author} {\bibfnamefont {G.~F.}\ \bibnamefont
  {CHAPLINE}},\ }\href {https://doi.org/10.1038/253251a0} {\bibfield  {journal}
  {\bibinfo  {journal} {Nature}\ }\textbf {\bibinfo {volume} {253}},\ \bibinfo
  {pages} {251} (\bibinfo {year} {1975})}\BibitemShut {NoStop}%
\bibitem [{\citenamefont {Gouttenoire}(2023)}]{gouttenoire_first-order_2023}%
  \BibitemOpen
  \bibfield  {author} {\bibinfo {author} {\bibfnamefont {Y.}~\bibnamefont
  {Gouttenoire}},\ }\href {https://doi.org/10.1103/PhysRevLett.131.171404}
  {\bibfield  {journal} {\bibinfo  {journal} {Physical Review Letters}\
  }\textbf {\bibinfo {volume} {131}},\ \bibinfo {pages} {171404} (\bibinfo
  {year} {2023})}\BibitemShut {NoStop}%
\end{thebibliography}%

\end{document}